# Binding Energies and Dissociation Temperatures of Heavy Quarkonia at Finite Temperature and Chemical Potential in the *N*-dimensional space

M. Abu-Shady[1], T. A. Abdel-Karim[1], E. M. Khokha[2]

Department of Applied Mathematics, Faculty of Science, Menoufia University, Shebin El- Kom,  Egypt[1]
Department of Basic Science, Modern Academy for Engineering and Technology, Cairo,  Egypt[2]

**Abstract**

The *N*-dimensional radial Schrödinger equation has been solved using the analytical exact iteration method (AEIM), in which the Cornell potential is generalized to finite temperature and chemical potential. The energy eigenvalues have been calculated in the *N*-dimensional space for any state (*n, l*). The present results have been applied for studying quarkonium properties such as charmonium and bottomonium masses. The binding energies and the mass spectra of heavy quarkonia are studied in the *N*-dimensional space. The dissociation temperatures for different states of heavy quarkonia are calculated in the three dimensional space. The influence of dimensionality number (*N*) has been discussed on the dissociation temperatures. A comparison is studied with other recent works. We conclude that the AEIM successes to predict the heavy-quarkonium properties at finite temperature and chemical potential.

## 1. Introduction

The solution of the radial Schrödinger equation with spherically symmetric potentials has vital applications in different fields of physics such as atoms, molecules, hadronic spectroscopy, and high energy physics. The Schrödinger equation has been solved by operator algebraic method [1], power series method [2-3], and path integral method [4]. Besides, Quasi-linearization method (QLM) [5], point canonical transformation (PCT) [6], Hill determinant method [7] and the conventional series solution method [8]. Recently, most of theoretical studies have developed interest in the solutions of radial Schrödinger equation in higher dimensions. These studies are more general and one can directly obtain the results in the lower dimensions.

The *N*-dimensional Schrödinger equation has been solved by various methods as the Nikiforov-Uvarov (UV) method [9-12], asymptotic iteration method (AIM) method [13], Laplace Transform method [14-15], (SUSQM) method [16], power series technique [17], Pekeris type approximation [18] and the analytical exact iteration method (AEIM) [19]. The *N*-dimensional radial Schrödinger equation has been solved for different types of spherical symmetric potentials as Coulomb potential [15], pseudoharmonic potential [20], Mie-Type potential [21], energy dependent potential [11] and Kartzer potential [22]. In addition, the Cornell potential type [13, 23] that consists of the Coulomb term and the linear term, an harmonic potential [14], the Cornell potential with harmonic oscillator potential [12] and the extended Cornell potential [19]. Additionally, the solution of Schrödinger equation at finite temperature has been used in different studies to describe the properties of heavy quarkonim systems.

Many efforts have been devoted to calculate the mass spectra of charmonuim and bottomonium mesons and to determine the binding energy and the dissociation temperatures of heavy quarkonia. In Refs. [24, 25], the authors have calculated the dissociation rates of quarkonium ground states by tunneling and direct thermal activation to the continuum and the binding energies and scattering phase shifts for the lowest eigenstates in the charmonium and bottomonium systems in hot gluon plasma. In Refs. [26, 27], the deconfinement and properties of the resulting quark–gluon plasma have been investigated by studying medium behavior of heavy quark bound states in statistical quantum chromodynamics and the spectral analysis of quarkonium states in a hot medium of deconfined quarks and gluons and the thermal properties of (QGP) are discussed. In Refs. [28-30], the authors have solved the Schrödinger equation at finite temperature for the charmonium and bottomonium states by employing an effective temperature dependent potential given by a linear combination of the color singlet free and internal energies and discussed the quarkonium spectral functions in a quark-gluon plasma. The dissociation of quarkonia has been studied by correcting the full Cornell potential through the hard-loop resumed gluon propagator, and HTL approximation [31, 32]. Moreover, the binding energies of the heavy quarkonia states are studied in detail in Refs. [33, 34].

At finite temperature and chemical potential, Vija and Thoma [35] have extended the effective perturbation theory for gauge theories at finite temperature and chemical potential for studying the collisional energy loss of heavy quarks in QGP. In Refs. [36, 37] have been generalized a thermodynamic quasi-particle description of deconfined matter to finite chemical potential and analyzed the response of color singlet and color averaged heavy quark free energies to a non-vanishing baryon chemical. On the same hand, the effect of chemical potential are studied on the photon production of QCD plasma, dissipative hydrodynamic effects on QGP, and thermodynamic properties of the QGP [38-42] by using different methods. At finite chemical potential and small temperature region the dissociation of quarkonia states have been studied in a deconfined medium of quarks and gluons in [43].

The aim of this work is to find the analytic solution of the $N$-dimensional radial Schrodinger equation with generalized Cornell potential at finite temperature and chemical potential using the exact analytical iteration method (EAIM) to obtain the energy eigenvalues. So far no attempt has been made to solve the $N$-dimensional radial Schrodinger when finite temperature and chemical potential are included. The application of present results on quarkonium properties such as the mass spectra of heavy quarkonium and the dissociation temperature for different states of heavy quarkonia have been calculated. Also, the influence of the dimensionality number has been investigated on the binding energy and the dissociation temperature at finite temperature and chemical potential.

The paper is organized as follows: In Sec. 2, the exact solution of the *N*-dimensional radial Schrödinger equation is derived. In Sec. 3, The binding energy and mass spectra of heavy quarkonia in the *N*-dimensional space are calculated. In Sec. 4, the dissociation temperature of heavy quarkonia in the *N*-dimensional space is obtained. In Sec. 5, Summary and conclusion are presented.

## 2. Exact Solution of the *N*-dimensional Radial Schrödinger Equation with the Cornell Potential at Finite Temperature and Chemical Potential.

The *N*-dimensional radial Schrödinger equation for two particles interacting via a spherically symmetric potential takes the form **[14, 44]**

$$\left[\frac{d^2}{dr^2} + \frac{N-1}{r}\frac{d}{dr} - \frac{l(l+N-2)}{r^2} + 2\mu_{Q\bar{Q}}(E_{nl} - U(r))\right]\psi(r) = 0, \tag{1}$$

where $l, N$ and $\mu_{Q\bar{Q}}$ are the angular quantum number, the dimensional number, and reduced mass of the two particles $\mu_{Q\bar{Q}} = \frac{m_Q m_{\bar{Q}}}{m_Q + m_{\bar{Q}}}$, respectively.

Now, Inserting $\psi(r) = R(r)/r^{(N-1)/2}$ in Eq. (1), we obtain

$$\left[\frac{d^2}{dr^2} - \frac{\lambda^2 - \frac{1}{4}}{r^2} + 2\mu_{Q\bar{Q}}(E_{nl} - U(r))\right]R(r) = 0, \tag{2}$$

with $\lambda = l + \frac{N-2}{2}$.

where $U(r)$ is the Cornell potential which take the form **[45]**

$$U(r) = \sigma r - \frac{\alpha}{r}. \tag{3}$$

With $\sigma = 0.192 \text{ GeV}^2$ and $\alpha = 0.471$. It is modified in QGP to study the binding energy and dissociation temperature by Debye screening mass as **[45-46]**

$$U(r, m_D) = \frac{\sigma}{m_D}\left(1 - e^{-m_D(T,\mu)r}\right) - \frac{\alpha}{r}e^{-m_D(T,\mu)r}, \tag{4}$$

where $m_D(T, \mu)$ is the Debye screening mass at finite temperature and quark chemical potential **[37]**, define as follows

$$m_D(T, \mu) = g(T)T\sqrt{\frac{N_c}{3} + \frac{N_f}{6}}\sqrt{1 + \frac{3N_f}{(2N_c + N_f)\pi^2}\left(\frac{\mu}{T}\right)^2}, \tag{5}$$

where $N_f$ is the number of quark flavor, $N_c$ the number of colors and $g(T)$ is the QCD coupling constant at finite temperature **[47]**.

$$g(T) = \frac{1}{(11N_c - 2N_f)\text{Log}(T^2/\Lambda_{QCD}^2)} \tag{6}$$

Using $e^{-m_D(T,\mu)r} = \sum_{k=0}^{\infty} \frac{(-m_D(T,\mu)r)^k}{k!}$ into Eq. (4) and neglect the higher orders when $m_D(T,\mu)r \ll 1$ and substituting into Eq. (4) reduce

$$U(r, m_D) = -ar^2 + br + c - \frac{d}{r}, \tag{7}$$

where,

$$a = \frac{1}{2}\sigma\, m_D(T,\mu),\ b = \frac{1}{2}(2\sigma - \alpha\, m_D(T,\mu)^2),\ c = \alpha\, m_D(T,\mu),\ d = \alpha. \tag{8}$$

Substituting from Eq. (7) into Eq. (2), we obtain

$$R''(r) = \left[-\varepsilon_{nl} + a_1 r^2 + b_1 r + c_1 - \frac{d_1}{r} + \frac{\lambda^2 - \frac{1}{4}}{r^2}\right] R(r), \tag{9}$$

where,

$$\varepsilon_{nl} = 2\mu_{Q\bar{Q}} E_{nl},\ a_1 = |-2\mu_{Q\bar{Q}} a|,\ b_1 = 2\mu_{Q\bar{Q}} b,\ c_1 = 2\mu_{Q\bar{Q}} c,\ d_1 = 2\mu_{Q\bar{Q}} d. \tag{10}$$

The analytical exact iteration method (AEIM) requires making the following ansatz for the wave function as in **[48-50]**.

$$R(r) = f_n(r)\, \exp[g_l(r)]. \tag{11}$$

Where

$$f_n(r) = \begin{cases} 1, & n = 0 \\ \prod_{i=1}^{n}(r - \alpha_i^{(n)}) & n = 1,2,3,\ldots \end{cases} \tag{12}$$

$$g_l(r) = -\frac{1}{2}\alpha r^2 - \beta r + \delta \ln r,\ \alpha > 0,\ \beta > 0. \tag{13}$$

From Eq. (11), we obtain

$$R''_{nl}(r) = \left(g''_l(r) + g'^2_l(r) + \frac{f''_n(r) + 2g'_l(r) f'_n(r)}{f_n(r)}\right) R_{nl}(r) \tag{14}$$

Comparing Eqs. (9) and (14) yields

$$a_1 r^2 + b_1 r + c_1 - \frac{d_1}{r} + \frac{\lambda^2 - \frac{1}{4}}{r^2} - \varepsilon = g''_l(r) + g'^2_l(r) + \frac{f''_n(r) + 2g'_l(r) f'_n(r)}{f_n(r)} \tag{15}$$

At ($n=0$), by substituting Eqs. (12) and (13) into Eq. (15) gives

$$a_1 r^2 + b_1 r + c_1 - \frac{d_1}{r} + \frac{\lambda^2 - \frac{1}{4}}{r^2} - \varepsilon_{0l}$$

$$= \alpha^2 r^2 + 2\alpha\beta r - \alpha[1 + 2(\delta + 0)] + \beta^2 - \frac{2\beta\delta}{r}$$

$$+ \frac{\delta(\delta - 1)}{r^2}. \tag{16}$$

By comparing the corresponding powers of $r$ on both sides of Eq. (16), one obtains

$$\alpha = \sqrt{a_1}, \tag{17a}$$

$$\beta = \frac{b_1}{2\sqrt{a_1}}, \tag{17b}$$

$$d_1 = 2\beta(\delta + 0), \tag{17c}$$

$$\delta(\delta + 1) = \lambda^2 - 1/4, \Rightarrow \delta = \frac{1}{2}(1 \pm 2\lambda), \tag{17d}$$

$$\varepsilon_{0l} = \alpha[1 + 2(\delta + 0)] + c_1 - \beta^2. \tag{17e}$$

From Eqs.(17a) – (17e), and (10). By taking the positive sign in Eq. (17d) then the ground state energy is:

$$E_{0l} = \sqrt{\frac{a}{2\mu_{Q\bar{Q}}}} (N + 2l) + c - \frac{b^2}{4a}. \tag{18}$$

For the first node ($n =1$), we use the functions $f_1(r) = \left(r - \alpha_1^{(1)}\right)$ and $g_l(r)$ from Eq. (8) to solve Eq. (10) gives:

$$a_1 r^2 + b_1 r + c_1 - \frac{d_1}{r} + \frac{\lambda^2 - 1/4}{r^2} - \varepsilon_{1l} =$$

$$\alpha^2 r^2 + 2\alpha\beta r - \alpha[1 + 2(\delta + 1)] + \beta^2 - \frac{2\left[\beta(\delta+1) + \alpha\alpha_1^{(1)}\right]}{r} + \frac{\delta(\delta-1)}{r^2} \tag{19}$$

Then, the relations between the potential parameters and the coefficients $\alpha, \beta, \delta$ and $\alpha_1^{(1)}$ are given by:

$$\alpha = \sqrt{a_1}, \tag{20a}$$

$$\beta = \frac{b_1}{2\sqrt{a_1}} \tag{20b}$$

$$d_1 = 2\beta(\delta + 1), \tag{20c}$$

$$\delta = \frac{1}{2}(1 \pm 2\lambda), \tag{20d}$$

$$\varepsilon_{1l} = \alpha[1 + 2(\delta + 1)] + c_1 - \beta^2, \tag{20e}$$

$$d_1 - 2\beta(\delta + 1) = 2\alpha \, \alpha_1^{(1)}, \tag{20f}$$

$$(d_1 - 2\beta\delta)\alpha_1^{(1)} = 2\delta, \tag{20g}$$

Substituting From Eqs. (12), into Eq. (4). We obtain the formula $E_{1l}$ as

$$E_{1l} = \sqrt{\frac{a}{2\mu_{Q\bar{Q}}}} (N + 2l + 2) + c - \frac{b^2}{4a}. \tag{21}$$

The second node ($n = 2$), we use $f_2(r) = \left(r - \alpha_1^{(2)}\right)\left(r - \alpha_2^{(2)}\right)$ and $g_l(r)$ from Eq. (13) to solve Eq. (15) gives:

$$a_1 r^2 + b_1 r + c_1 - \frac{d_1}{r} + \frac{\lambda^2 - \frac{1}{4}}{r^2} - \varepsilon_{2l}$$

$$= \alpha^2 r^2 + 2\alpha\beta r - \alpha[1 + 2(\delta + 2)] + \beta^2 - \frac{2\left[\beta(\delta + 2) + \alpha\left(\alpha_1^{(2)} + \alpha_2^{(2)}\right)\right]}{r}$$

$$+ \frac{\delta(\delta - 1)}{r^2}. \tag{22}$$

Thus, the relations between the coefficients $\alpha, \beta, \delta, \alpha_1^{(2)}$ and $\alpha_2^{(2)}$ are given by:

$$\alpha = \sqrt{a_1}, \tag{23a}$$

$$\beta = \frac{b_1}{2\sqrt{a_1}}, \tag{23b}$$

$$\delta = \frac{1}{2}(1 \pm 2\lambda), \tag{23c}$$

$$\varepsilon_{2l} = \alpha[1 + 2(\delta + 2)] + c_1 - \beta^2, \tag{23d}$$

$$d_1 - 2\beta(\delta + 2) = 2\alpha\left(\alpha_1^{(2)} + \alpha_2^{(2)}\right), \tag{23e}$$

$$(d_1 - 2\beta\delta)\alpha_1^{(2)}\alpha_2^{(2)} = 2\delta(\alpha_1^{(2)} + \alpha_2^{(2)}), \tag{23f}$$

$$[d_1 - 2\beta(\delta + 1)]\left(\alpha_1^{(2)} + \alpha_2^{(2)}\right) = 4\alpha\,(\alpha_1^{(2)}\alpha_2^{(2)}) + 2(2\delta + 1). \tag{23g}$$

Hence, the formula $E_{2l}$ is given by

$$E_{2l} = \sqrt{\frac{a}{2\mu_{Q\bar{Q}}}}(N + 2l + 4) + c - \frac{b^2}{4a}. \tag{24}$$

Then, the iteration method is repeated many times. Therefore, the exact energy formula for any state at finite temperature and chemical potential in *N*-dimensional space is written as:

$$E_{nl}^N = \sqrt{\frac{a}{2\mu_{Q\bar{Q}}}}(N + 2l + 2n) + c - \frac{b^2}{4a}, \quad n = 0,1,2,\dots \tag{25}$$

### 3- Discussion of Results

In the first, We compare between the exact potential U(r,T,µ) and the approximate potential V(r,T,µ) for different vaules of chemical potential and temperature.

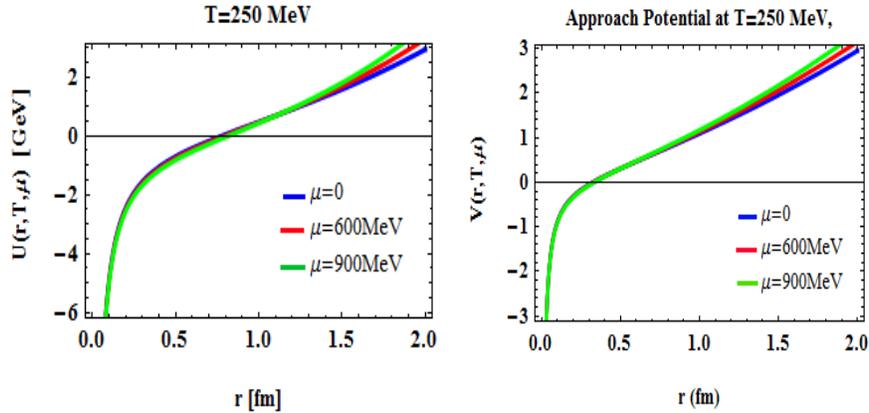

**Fig. 1.** The Cornell potential at finite temperature and chemical potential as a function of a distance ($r$) for different chemical potential. The exact potential $U(r,T,\mu)$ in the left panel and the approximate potential $V(r,T,\mu)$ in the right panel.

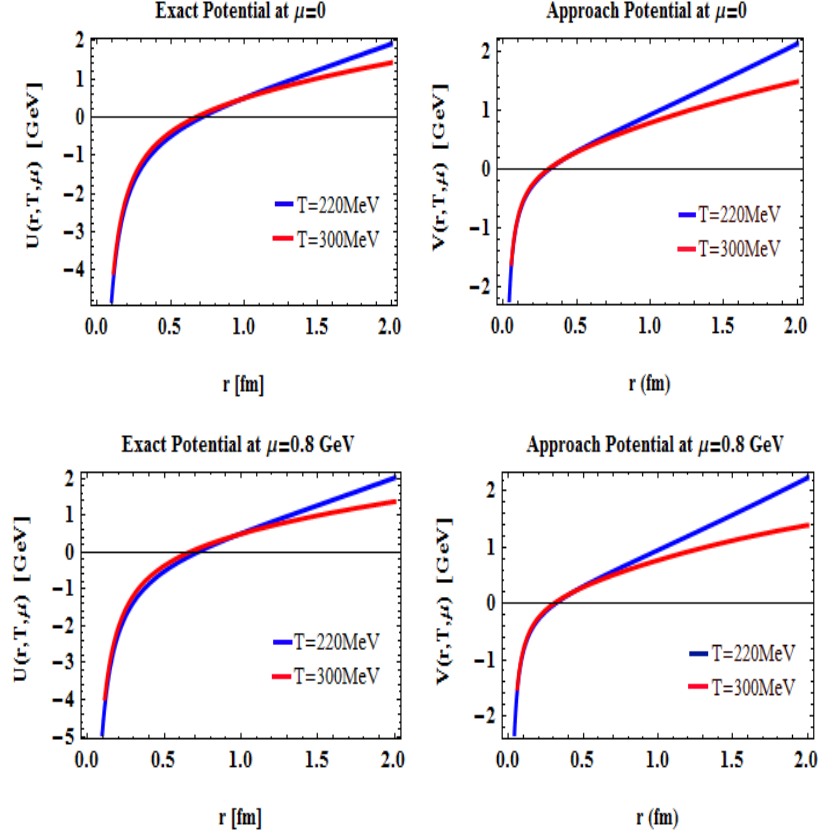

**Fig. 2.** The Cornell potential at finite temperature and chemical potential as a function of a distance (*r*).for different temperature. The exact potential $U(r,T,\mu)$ in the left panels and the approach potential $V(r,T,\mu)$ in the right panels.

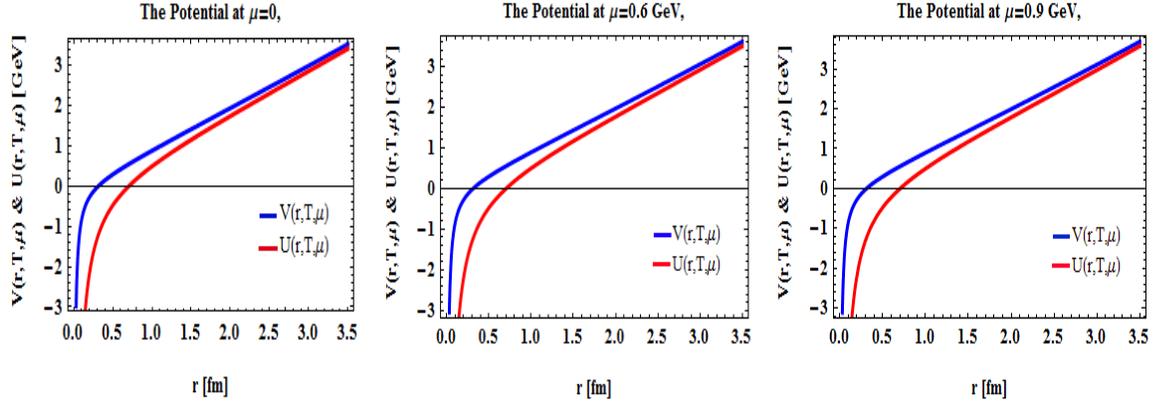

**Fig. 3.** The comparison between the different potentials. The red line is the exact potential $U(r,T,\mu)$, the blue line is the approxmite potential $V(r,T,\mu)$ for different chemical potential.

In **Fig.1**, both of the exact $U(r,T,\mu)$ and the approach potential $V(r,T,\mu)$ are plotted for different chemical potentials. We note a good qualitative agreement between exact potential and approximate potential, in which an increase in chemical potential leads to decreasing in the positive part of the two potential. Also, when chemical potential is fixed, we note that the qualitative agreement is observed between exact potential and approximated potential. This behavior is in agreement with Refs. **[51-53]**. In **Fig. 2.** The comparison between the exact potential $U(r,T,\mu)$ and the approximate potential $V(r,T,\mu)$ is plotted as a function of distance $r$ for different values of chemical potentials. The results in **Fig. 2.** shows that the behavior of the two potentials are similar.

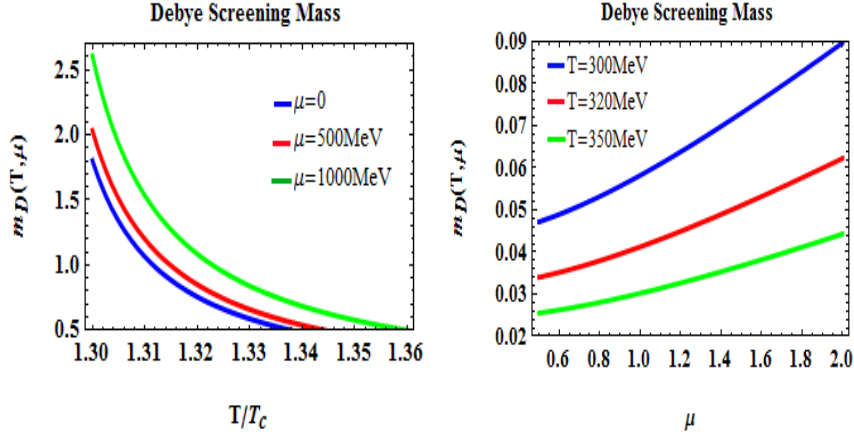

**Fig. 3.** Debye screening mass with temperature at different values of chemical potential (left panel) and Debye screening mass with chemical potential at different values of temperatures (right panel).

In **Fig. 3**.The Debye screening mass is plotted with temperature at different values of chemical potential (left panel) and with chemical potential at various temperatures (right panel). The left panel shows that the Debye screening mass decreases with the temperature but shifts to upper values by increasing chemical potential that is in agreement with Refs. **[54, 55]**. The right panel shows that the Debye screening mass increases with the chemical potential but shifts to lower values by increasing the temperature in agreement with Ref. **[43]**.

## 3. Binding Energy and Mass Spectra of Heavy Quarkonia in *N*-dimensional space

In this section, the binding energy and the mass spectra of heavy quarkonia are calculated such as charmonium and bottomonium in the *N*-dimensional space for any state at finite temperature and chemical potential.

Substituting from Eq. (8) into Eq. (25). Therefore, the binding energies for the different states of heavy quarkonia at finite temperature and chemical potential take the form

$$E_{bin}(T,\mu) = \alpha\, m_D(T,\mu) + \sqrt{\frac{\sigma\, m_D(T,\mu)}{4\, \mu_{Q\bar{Q}}}}(2n + 2l + N) - \frac{\left(2\sigma - \alpha\, m_D^{\,2}(T,\mu)\right)^2}{8\, \sigma\, m_D(T,\mu)}, \qquad (26)$$

where $\mu_{Q\bar{Q}} = m_c/2$ for charmonium meson and $\mu_{Q\bar{Q}} = m_b/2$ for bottomonium meson. The behavior of the binding energy for the different states of heavy quarkonia is shown in **Figs. 4, 5 and 6.**

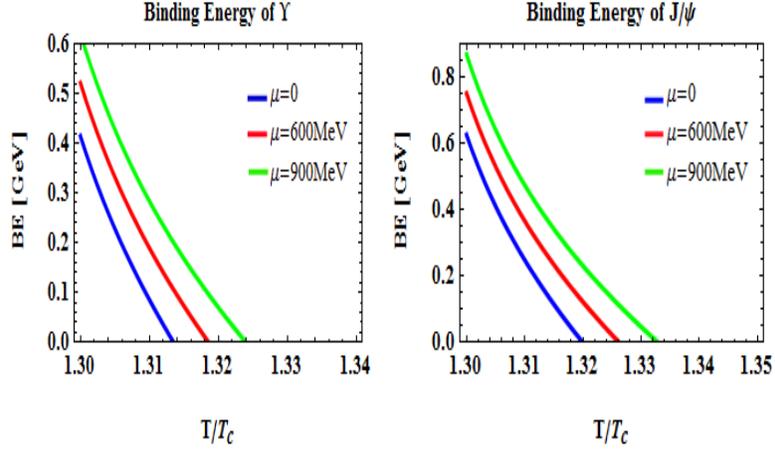

**Fig. 4.** Dependence of ϒ binding energy (in GeV) on temperature T / T$_c$ (left panel) and dependence of J/ψ binding energy (in GeV) on temperature T / T$_c$ (right panel) at different values of chemical potential.

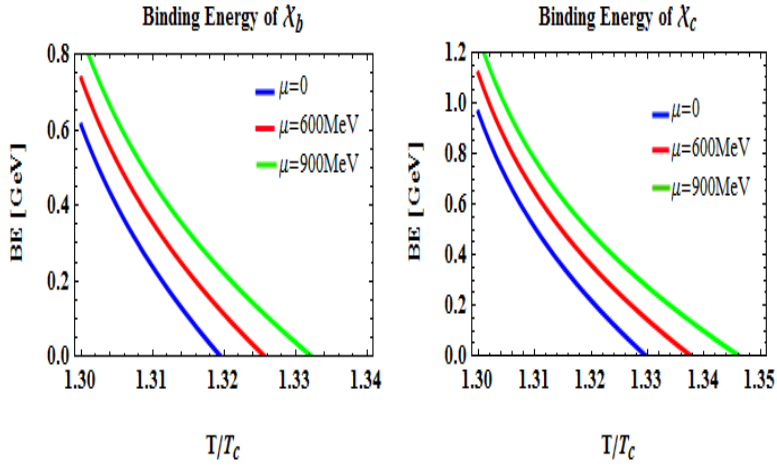

**Fig. 5.** Dependence of χ$_b$ binding energy (in GeV) on temperature T / T$_c$ (left panel) and dependence of χ$_c$ binding energy (in GeV) on temperature T / T$_c$ (right panel) at different values of chemical potential.

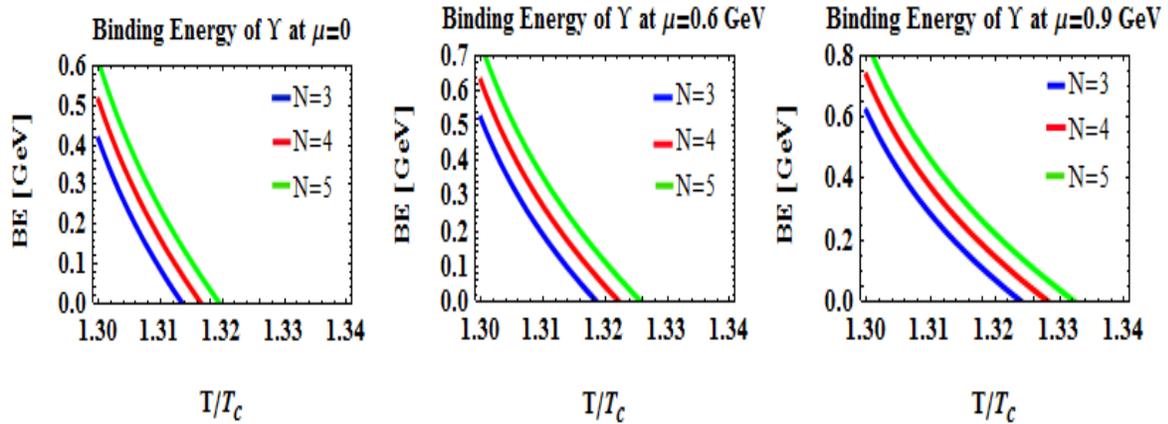

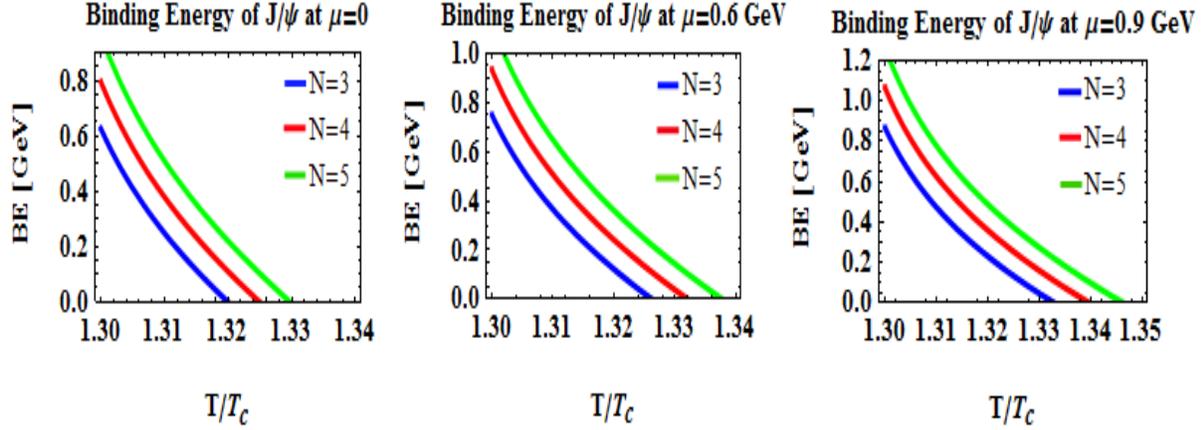

**Fig. 6.** Dependence of ϒ binding energy (in GeV) on temperature T / $T_c$ ( upper panels ) and dependence of J/ψ binding energy (in GeV) on temperature T / $T_c$ ( lower panels ) at different values of N.

**Figs. (4, 5)** show the behavior of the binding energy of heavy quarkonia as a function of temperature (in units of $T_c$) for 1S and 1P states respectively. Clearly, we observed that the binding energy becomes weaker with increasing temperature. The dependence of the binding energy on the temperature shows a qualitative agreement with similar results in. **[25, 29- 33]**, and becomes stronger with the chemical potential. **Fig. 6**, shows the dependence of the binding energy of J/ψ and ϒ states on dimensionality number (N). The binding energy of $\frac{J}{\psi}$ and ϒ states increases with the increasing dimensionality number. In **Fig. 7**. The binding energy of charmonium and bottomonium mesons is plotted in the 3- dimensional space. We note that the binding energy increases with increasing finite temperature and chemical potential. The effect of finite temperature is more effect than the chemical potential.

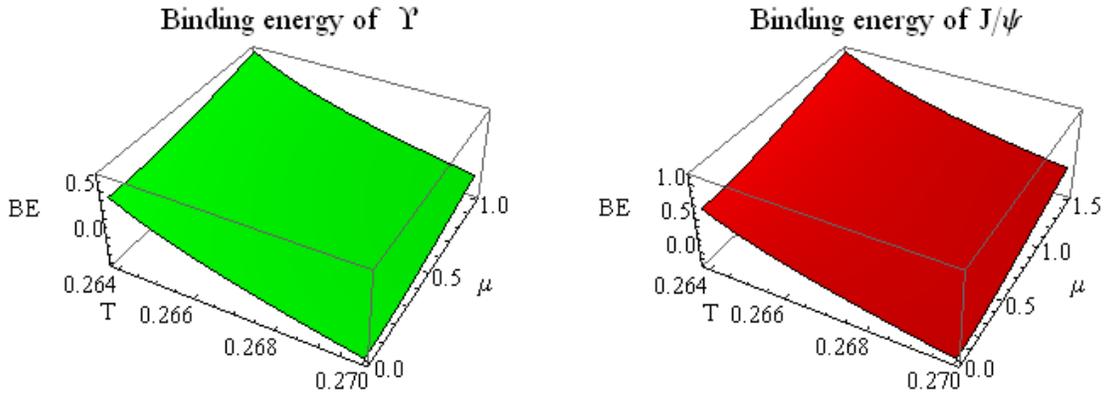

**Fig. 7.** Dependence of ϒ binding energy (in GeV) on temperature T / $T_c$ (left panel) and dependence of J/ψ binding energy (in GeV) on temperature T / $T_c$ (right panel) in 3- dimensions.

For calculating the mass spectra of heavy quarkonia, the following relation is used **[13]**.

$$M = 2\, m_Q + E_{nl}^N . \tag{27}$$

Substituting from Eq. (26) into Eq. (27). Thus, the mass spectra for the different states of heavy quarkonia as a function of temperature and chemical potential take the following form:

$$M(T,\mu) = 2\, m_Q + \alpha\, m_D(T,\mu) + \sqrt{\frac{\sigma\, m_D(T,\mu)}{4\, \mu_{Q\bar{Q}}}(2n + 2l + N)} - \frac{(2\sigma - \alpha\, m_D^2(T,\mu))^2}{8\, \sigma\, m_D(T,\mu)}. \quad (28)$$

In **Fig. 8**, the mass spectra of heavy quarkonia is plotted as a function of temperature for **1S** and **1P** states bottomonium in the upper panels and charmonuim in the lower panels. We note that the mass spectra decreases with increasing temperaturesf for 1S and 1P states. The values of 1P state is larger than the values of 1S state for different values of chemical potential. Thus, the finite temperature and chemical potential act increasingly on the 1P state as well as 1S,

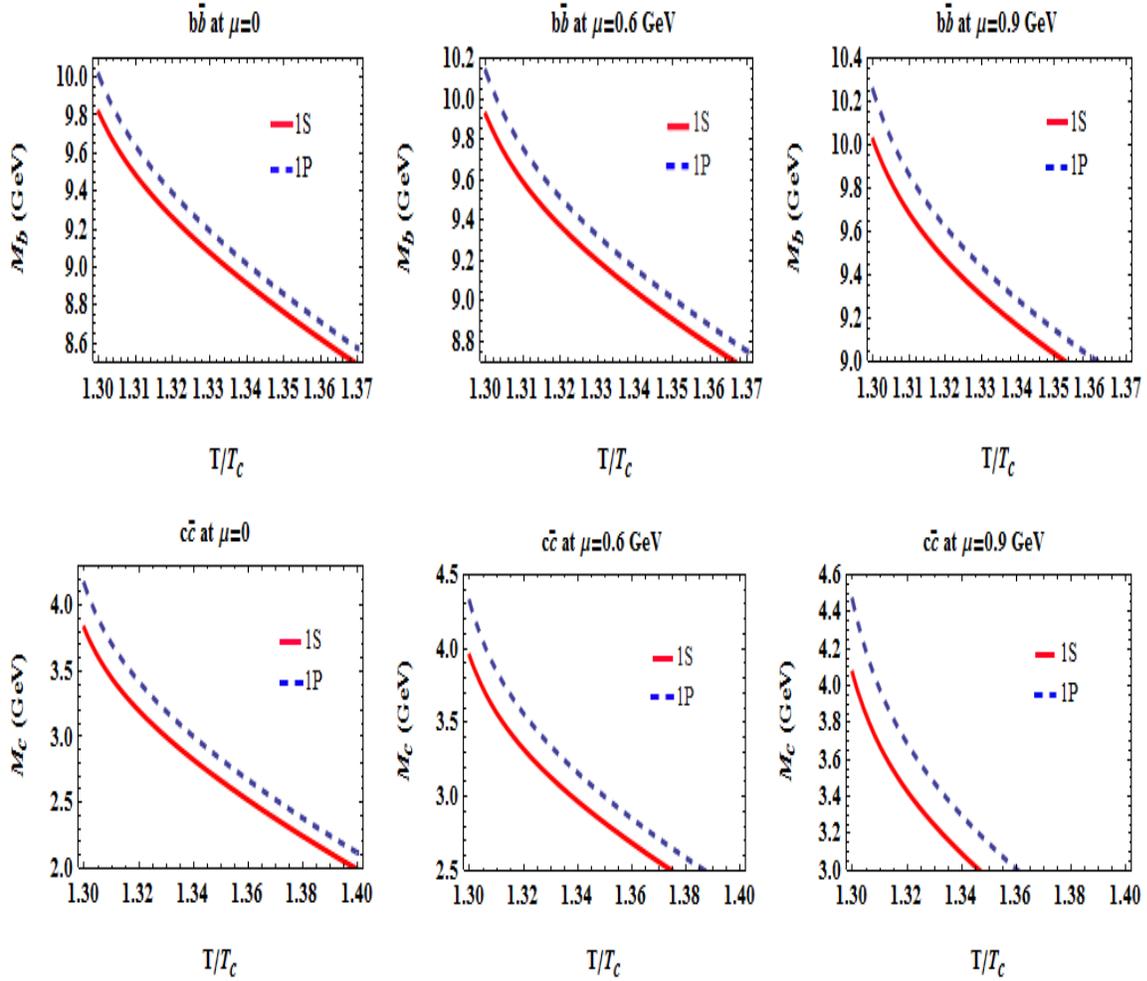

Fig. 8. The mass spectra of heavy quarkonia as a function of temperature for **1S** and **1P** states, bottomonium in the upper panels and charmonuim in the lower panels

In **Figs. (9, 10)**, the behavior of the mass spectra of heavy quarkonia is studied as a function of temperature (in units of $T_c$) for 1S and 1P states at various chemical potential using two values of quark mass. An increase of quark mass leads to increasing mass spectra of heavy quarkonia for states 1S and 1P that agreement with Ref. [28].

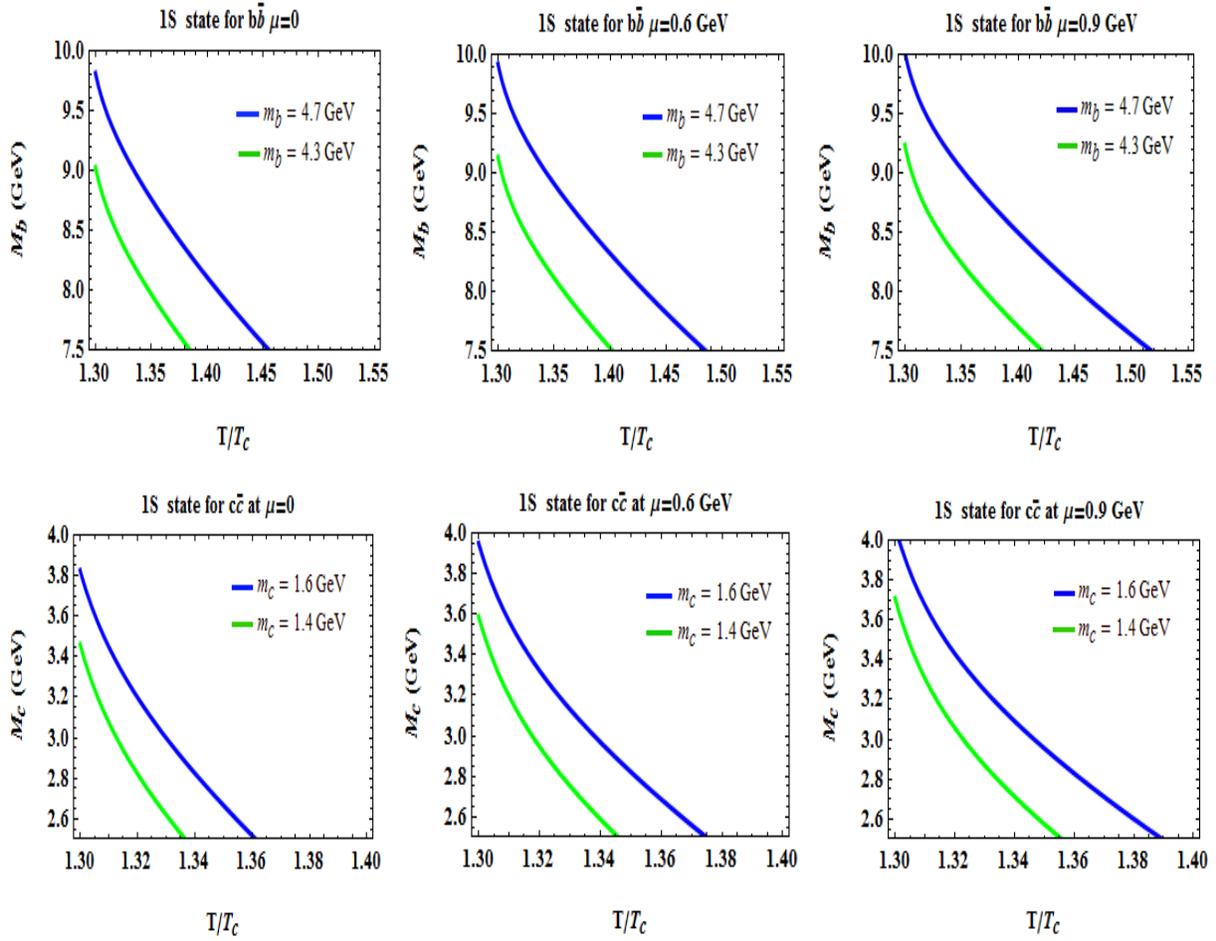

**Fig. 9.** The mass spectra of heavy quarkonia as a function of temperature for **1S** states, bottomonium in the upper panels and charmonuim in the lower panels.

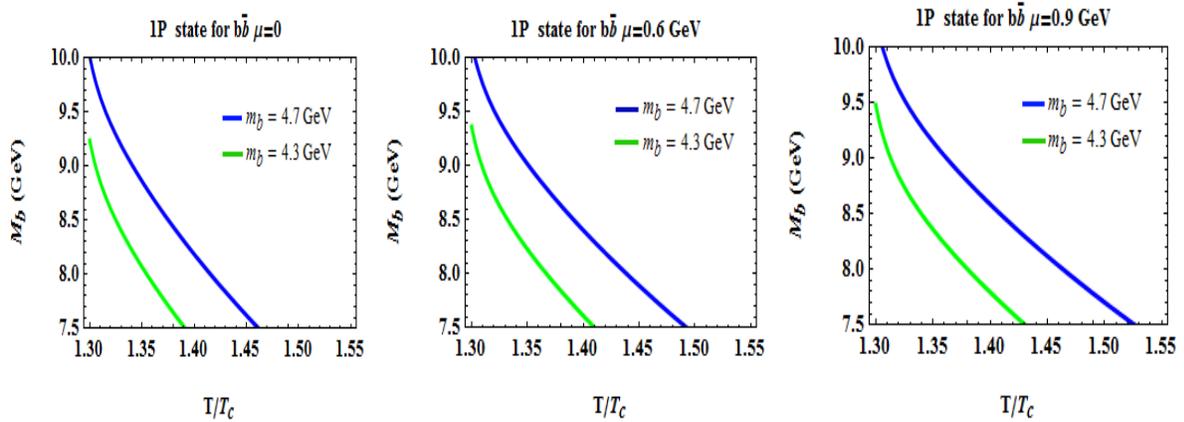

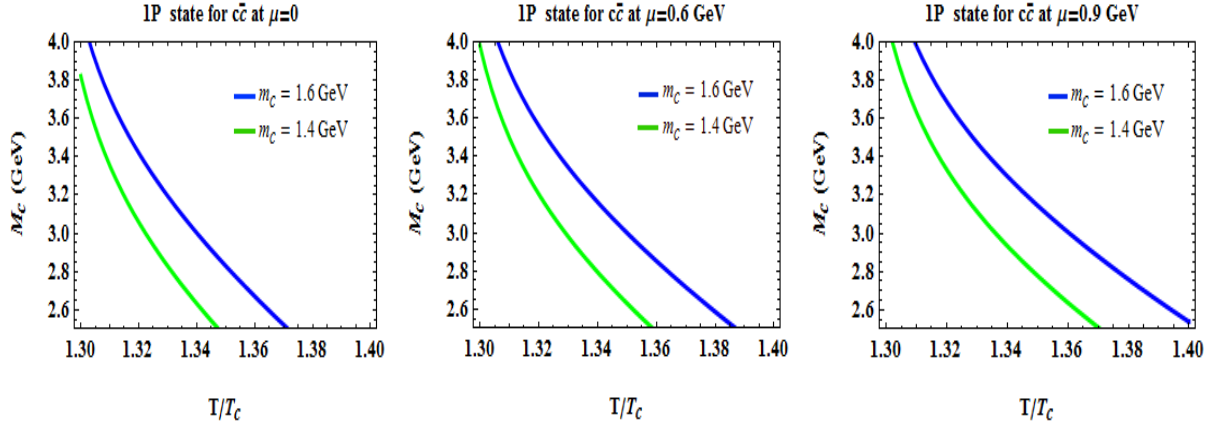

**Fig. 10.** The mass spectra of heavy quarkonia as a function of temperature for **1P** state, bottomonium in the upper panels and charmonuim in the right panels

## 4. Dissociation Temperature of Heavy Quarkonia in *N*-dimensional space

There are a lot of earlier studies for determining the dissociation temperatures for different states of heavy quarkonia. In Ref. **[24],** authors have calculated the dissociation temperature of the heavy quarkonia from the thermal width $\Gamma(T)$. In Ref. **[29]**, authors have put a conservative condition for the dissociation $\Gamma(T) > 2E_{bin}$. In Ref. [30], the authors have calculated the upper bound and the lower bound of the dissociation temperature ($T_D$) by the condition for the dissociation: $E_{bin} = T_D$ and $E_{bin} = 3T_D$ respectively. In Ref. [43], the authors have obtained the dissociation temperature of quarkonia when their binding energies are of the order of the baryon chemical potential.

we can calculate the dissociation temperature for different states of heavy quarkonia from the condition: $E_{bin} = 0$. Since, the state is dissociated when its binding energy vanished as **[27]**.

**Table. 1.** The dissociation temperature ($T_D$) with $T_c$ = 203 MeV for the quarkonia states (in units of $T_c$) using $m_c$ = 1.6 GeV and $m_b$ = 4.7 GeV at $\mu = 0$.

| State | N=3 | N=4 | N=5 |
|---|---|---|---|
| **J/ψ** | 1.31997$T_c$ | 1.32493 $T_c$ | 1.32974 $T_c$ |
| ψ´ | 1.32974$T_c$ | 1.33441 $T_c$ | 1.33897 $T_c$ |
| ϒ | 1.31351$T_c$ | 1.31656 $T_c$ | 1.31955 $T_c$ |
| ϒ´´ | 1.31955$T_c$ | 1.32247 $T_c$ | 1.32534 $T_c$ |

In **Table 1.** We have calculated the dissociation temperature for the ground state and the first excited states of $\bar{c}c \text{ and } \bar{b}b$ at $N = 3$ and also at higher dimensional space N = 4 and N = 5 when chemical potential is vanished. It is noted from Table 1 that the states are dissociated around 1.3 $T_c$. The values of J/ψ and ϒ´ are agree quantitatively with the recent reported Agotiya et al. [30]. ϒ gives smaller value in comparison with Ref. [30] which equals 1.7 $T_c$. Also, the value ϒ are agreement with Ref. [56] which gives the dissociation temperature of the 1S bottomonium $T_{d=}1.4\ T_C$. In Ref. [57], the dissociation temperature depends on the chosen of the Debye screening mass. It important to display the effect of dimensionality number on the dissociation temperature. It note of Table 1 that the increasing dimensionality number leads to increasing the dissociation temperature at zero chemical potential.

**Table. 2.** The dissociation temperature $T_D$ (MeV) at $\mu = 0.6$ GeV with $T_c = 185$ MeV.

| State | $N=3$ | $N=4$ | $N=5$ |
|---|---|---|---|
| **J/ψ** | 1.45524 $T_c$ | 1.46167 $T_c$ | 1.46789 $T_c$ |
| ψ' | 1.46789 $T_c$ | 1.47393 $T_c$ | 1.47982 $T_c$ |
| Υ | 1.44687 $T_c$ | 1.45082 $T_c$ | 1.45469 $T_c$ |
| Υ' | 1.45469 $T_c$ | 1.45848 $T_c$ | 1.46219 $T_c$ |

**Table. 3.** The dissociation temperature $T_D$ (MeV) at $\mu = 0.9$ GeV with $T_c = 160$ MeV.

| State | $N=3$ | $N=4$ | $N=5$ |
|---|---|---|---|
| **J/ψ** | 1.69108 $T_c$ | 1.69973 $T_c$ | 1.70811 $T_c$ |
| ψ' | 1.70811 $T_c$ | 1.71624 $T_c$ | 1.72417 $T_c$ |
| Υ | 1.67982 $T_c$ | 1.68514 $T_c$ | 1.69035 $T_c$ |
| Υ' | 1.69035 $T_c$ | 1.69545 $T_c$ | 1.70045 $T_c$ |

In **Tables 2** and **3**. The dissociation temperatures for different states of heavy quarkonia have been obtained at finite chemical potential, where the critical temperatures ($T_c$) are taken as 185 MeV and 160 MeV for $\mu = 600$ MeV and $\mu = 900$ MeV, respectively as in Ref. **[58]**. There is an important observation, an increase in the value of chemical potential increases the value of dissociation temperatures. Also, one notes that increasing dimensional number leads to a small increase in the dissociation temperatures. Therefore, the finite temperature and dimensional number play an important role in changing dissociation temperatures which are not taken in account in many previous works.

## 5. Summery and Conclusion

In this paper, we have employed the analytical exact iteration method (AEIM) for determining the exact solution of the *N*-dimensional radial Schrödinger equation, in which the Cornell potential is generalized at finite temperature and chemical potential. The energy eigenvalues have been calculated in the *N*-dimensional space for any state (*n, l*), in which one can obtain the energy eigenvalues in lower dimensions in agreement with recent works. The present results are applied on studying the properties of heavy quarkonia such as charmonium and bottomonium.

The effect of temperature, chemical potential, and dimensionality number is studied on the binding energies and the mass spectra of heavy quarkonia. The present results are in agreement with recet works [25, 29, 33]. The binding energies of 1S and 1P states for charmonium and bottomonium with the temperature have been studied in comparison with other studies [11, 30]. Additionally, the effect of the dimensionality number ($N$) on the values in the dissociation temperatures of heavy quarkonia at zero and finite chemical potential has been studied.

The novelty in this work that we have found the exact solution of the $N$-dimensional radial Schrödinger equation with the Cornell potential at finite temperature and chemical potential using the analytical exact iteration method (AEIM). The binding energies and the mass spectra of heavy quarkonia have been investigated in the $N$-dimensional space at finite temperature and chemical potential. We have determined the dissociation temperatures for different states of heavy quarkonia at finite temperature and chemical potential. The present results show a qualitative agreement with the earlier studies.

## 6- References